\documentclass[12pt]{article}

\usepackage[a4paper,left=2cm,right=2cm,top=1cm,
bottom=2cm]{geometry}

\usepackage{times}
\usepackage{amsmath}
\usepackage{siunitx}
\usepackage[T1]{fontenc}
\usepackage[english]{babel}
\usepackage[latin2]{inputenc}  
\usepackage{epsfig}
\usepackage{epstopdf}
\usepackage{multirow}
\usepackage{graphicx}
\usepackage{caption}
\usepackage{url}
\usepackage{float}
\usepackage{array}
\usepackage{multirow}
\usepackage{amssymb}
\usepackage{caption}
\usepackage{subcaption}
\usepackage{titlesec}
\usepackage{cite}
\usepackage[table]{xcolor}
\usepackage{adjustbox} 
\usepackage{lipsum}
\usepackage[labelfont=bf]{caption}
\usepackage{color}

\usepackage[deletedmarkup=xout]{changes}
\definechangesauthor[color=blue]{SW}
\definechangesauthor[color=red]{TL}

\newenvironment{sciabstract}{%
\begin{quote} }
{\end{quote}}

\newcounter{lastnote}

\title{ Electron power absorption in low pressure  capacitively coupled electronegative oxygen radio frequency plasmas }

\author
{M\'at\'e Vass$^{1\ast}$, Sebastian Wilczek$^{2}$,  Trevor Lafleur$^{4}$,\\ Ralf Peter Brinkmann$^{2}$, Zolt\'an Donk\'o$^{1}$, Julian Schulze$^{2,3}$ \\
}

\date{}

\begin{document}

\maketitle 
 \vspace{-1cm}
 {\small
\begin{flushleft}

 $^{1}$  Institute for Solid State Physics and Optics, Wigner Research Centre for Physics, H-1121 Budapest, Konkoly-Thege Mikl\'os str. 29-33, Hungary\\

 $^{2}$  Department of Electrical Engineering and Information Science, Ruhr-University Bochum, D-44780, Bochum, Germany \\
 
$^{3}$ Department of Physics, West Virginia University, Morgantown, WV 26506, United States of America \\

$^{4}$ PlasmaPotential--Physics Consulting and Research, Canberra, ACT 2601, Australia \\
\vspace{0.5cm}
E-mail: vass.mate@wigner.mta.hu

\end{flushleft}
}
\vspace{0.3cm}

\begin{sciabstract}
 A thorough understanding of the energy transfer mechanism from the electric field to electrons is of utmost importance for optimization and control of different plasma sources and  processes. This mechanism, called electron power absorption, involves complex electron dynamics in electronegative capacitively coupled plasmas (CCPs) at low pressures, that are still not fully understood. Therefore, we present a spatio-temporally resolved
analysis of electron power absorption in low pressure oxygen CCPs based on the momentum balance equation derived from the Boltzmann equation. Data are obtained from 1d3v Particle-In-Cell / Monte Carlo Collision simulations. In contrast to conventional theoretical models, which predict  `stochastic/collisionless heating' to be important at low pressure, we observe the dominance of Ohmic power absorption. In addition, there is an attenuation of ambipolar power absorption at low pressures due to the strong electronegativity, and the presence of electropositive edge regions in the discharge, which cause a high degree of temporal symmetry of the electron temperature within the RF period.

\end{sciabstract}

\section{Introduction}

Capacitively coupled radio frequency (RF) plasma sources are of great relevance due to their numerous industrial applications \cite{LibermanBook, ChabertBook, MakabeBook}. Because of their complexity, plasma sources exhibit a wide range of physical mechanisms, some of which are still not fully understood, although their complete understanding would immensely contribute to the optimisation process of such industrial applications. One of the most important issues is the electric field-electron energy transfer mechanism, which is vital for generating and sustaining the plasma \cite{schulze16}. This mechanism is commonly called `electron heating'. However, this term is problematic because what is meant is the total transfer of energy into the plasma, not just the fraction related to the increase in electron temperature. Therefore, `electron power absorption' is used instead \cite{schulze18}.

There have been numerous attempts to model the electron power absorption. A periodically varying electric field can only deposit energy to a charged particle on time average provided the phase coherence between the electric field and the particle motion is broken by some process. A reasonable candidate for such a process is collisions  between electrons and gas atoms/molecules. This process is commonly called  `Ohmic heating', and is modeled by a simplified version of the first velocity moment of the Boltzmann equation (i.e. the momentum balance equation), where gradients are neglected, a harmonic time dependence of the electric field is assumed, and a simplified version of the collisional electron momentum loss is used \cite{LibermanBook, ChabertBook, MakabeBook}. Godyak {\it et al.} \cite{Godyak1985,Godyak1990} were the first to show that this description is incomplete and there has to be an additional form of heating mechanism sustaining low pressure CCPs, called  `collisionless heating', which is not incorporated in this simplified framework. A separate model, called the Hard Wall Model was developed to describe this `collisionless' (or `stochastic') electron power absorption \cite{Godyak1972,Lieberman1988,Lafleur2014x,Turner2009}. This model uses the following assumptions: (i) The electron density is zero inside the sheath and changes abruptly at the sheath edge.  (ii) The electrons are accelerated during sheath expansion by the space- and time-dependent electric field, which is assumed to be high inside the sheath and zero outside. Furthermore, the model treats only a single sheath and assumes an input electron energy distribution function. Hence, the electron-field phase interaction is not correctly modelled, and the assumptions are not of general validity, as the electric field does not vanish outside the sheath, and is not a harmonic function of time in electropositive single-frequency low-pressure CCPs \cite{schulze15,schulze18}. The total electron power absorption is calculated as the sum of the `collisionless heating' and the  `Ohmic heating'. It is also important to stress, that these models are conceptually separate and hence do not give a self-consistent description of the problem, although in some cases their predictions are in reasonable agreement with experiment \cite{Godyak1985,Godyak1990}. 

During the investigation of electron power absorption in low pressure capacitively coupled plasmas, several new phenomena were observed, such as the `nonlinear electron resonance heating' (NERH) \cite{MussenbrockNERH,Wilczek2} and the importance of plasma-surface interactions, which  led to the introduction of different modes of discharge operations, such as the $\alpha-$, $\gamma-$, DA- (drift-ambipolar) and striation modes \cite{Belenguer1990,schulze2011,Czarnetzki1999,schulze2008,schuengel2013,Liu2015, Gudmundsson2017,gud, gud2018,gud2019,derzsi, Huang2015,Liu2016, Liu2017}.

Surendra and Dalvie were the first to analyse the electron power absorption self-consistently using the Boltzmann equation and PIC-simulations in an electropositive single-frequency low-pressure CCP \cite{Surendra1993}. They found that the electron pressure gradient term, originally neglected in the model for  `Ohmic heating', has a strong contribution to `collisionless heating'. Furthermore, they concluded that the space- and time-dependence of the mean electron energy in CCPs is important to obtain a nonzero  `collisionless heating' on time average. Turner and Gozadinos developed a theory of  `pressure heating' in CCPs based on similar arguments \cite{turner, Gozadinos}. Lafleur {\it et al.} revisited electron power absorption with a similar approach as Surendra and Dalvie \cite{lafleur14}. They found that only `Ohmic heating' and `pressure heating' have significant contributions to the total electron power absorption on time average, although, under certain conditions, inertial power absorption can have a nonzero contribution on time average. Based on his Smooth Step Model (SSM), Brinkmann derived a unified description of electron power absorption in CCPs \cite{Brinkmann2015, Brinkmann2016}.  He showed that the total electron power absorption on time average is the sum of four terms, each one corresponding to one of the power absorption mechanisms known from separate previous theories, i.e.  NERH,  `stochastic heating' (releated to the Hard Wall Model), `ambipolar/pressure heating', and `Ohmic heating'.  He also demonstrated that a time dependence of the electron temperature is necessary to obtain a non-zero `collisionless electron heating' on time average \cite{Brinkmann2015, Brinkmann2016}. Grapperhaus and Kushner developed a semianalytic RF sheath model to describe electron power absorption in CCPs \cite{Grapperhaus}. 

More recently, Schulze {\it et al.} performed a detailed  space- and time-resolved analysis of the electron power absorption dynamics in electropositive single frequency argon CCPs based on the first two moment equations of the Boltzmann equation \cite{schulze18}. They showed that the electric field is high outside the sheath, mainly due to the ambipolar electric field ( $E_{\nabla n}=-\frac{T_{xx}}{ne}\frac{\partial n}{\partial x}$, where $T_{xx}$ is the electron temperature and $n$ is the electron density) and that the sheath expansion and collapse phases are not `mirror' images of each other. They provided theoretical evidence, that a constant electron temperature (the isothermal assumption, present in some models for electron power absorption) would lead to zero power absorption generated by the ambipolar electric field on time average, which is the most important term at low pressures in electropositive single frequency argon CCPs. They explained the temporal asymmetry of the electron temperature, which is intimately related to the ambipolar electric field as follows: During sheath expansion electrons are accelerated by the ambipolar electric field towards the bulk. As the electrons gain energy from the ambipolar electric field, the electron temperature increases. Due to the fact, that the ambipolar electric field is proportional to the electron temperature, its increase leads to an increase of the ambipolar electric field as well, thus leading to a self-amplifying mechanism, which stops when the sheath is fully expanded. The same mechanism does not  operate during sheath collapse, as in this case cold electrons move towards the electrode, while the direction of $E_{\nabla n}$ remains the same, since it is determined by the ions, which generally do not react to the RF excitation. As the direction of the electron current is reversed compared to the sheath expansion phase, electrons are cooled by $E_{\nabla n}$ during sheath collapse and $T_{xx}$ is low. Furthermore, it was shown, that the gradient of the electron temperature has an important contribution to the electron power absorption dynamics. It was also pointed out that  Ohmic  power absorption increases as a function of pressure, whereas pressure heating (the sum of ambipolar  power absorption and the power absorption originating from the temperature gradient, see next section) remains nearly constant at all pressures and, thus, is less significant at higher pressures due to the dominance of the Ohmic power absorption term.

In this work, we use the same analysis as in \cite{schulze18} for single-frequency low pressure electronegative oxygen CCPs. There has been a growing interest in oxygen CCP discharges \cite{Gudmundsson2017, gud, gud2018, gud2019, derzsi, greb1, greb2, proto, aranka, Zhang, Dittmann, Dittmann2, Gibson, Matthias, Wang, You, Lisovsky}.  The electron power absorption dynamics in oxygen has been studied in \cite{ Gudmundsson2017, gud, gud2018, gud2019, derzsi, Lisovsky}. Lisovsky {\it et al.} investigated $\alpha$-$\gamma$ mode transitions in low pressure oxygen CCPs \cite{Lisovsky}. Gudmundsson {\it et al.} investigated single frequency oxygen discharges at low pressure. They concluded, that the discharge undergoes a  power absorption mode transition from a hybrid DA-$\alpha$ mode at low pressure to a pure $\alpha$-mode at higher pressure \cite{Gudmundsson2017, gud}. They also investigated mode transitions at constant pressure by changing the electrode gap and the driving frequency. It was found that by increasing the driving frequency the electronegativity of the oxygen discharge decreases and thus a power absorption mode transition can occur \cite{gud2018}. Furthermore, at constant pressure and driving frequency, the smaller the gap, the more electronegative the discharge is due to the increased significance of surface quenching. \cite{gud2019}. Derzsi {\it et al.} investigated electron power absorption mode transitions in oxygen discharges driven by tailored voltage waveforms \cite{derzsi}. They concluded that changing the number of consecutive harmonics included in the driving voltage waveform has a strong effect on the electronegativity of the discharge. At higher base frequencies, increasing the number of harmonics results in the decrease of the electronegativity, whereas at low base frequencies this has no effect.

We show that whereas at high pressures, where the electronegativity of oxygen is low \cite{ang}, the electron power absorption dynamics is similar to that of argon, at low pressues, where oxygen is highly electronegative, a different phenomenon occurs: The total contribution of  the power absorption term associated with the pressure gradient becomes insignificant and Ohmic  power absorption will be the most pronounced term in the total electron power absorption, despite the low neutral gas pressure. This is in marked contrast to expectations, which claim that `collisionless heating' dominates over Ohmic  power absorption at low pressures. We conclude that the attenuation of the ambipolar field is linked to a temporally symmetric electron temperature distribution, which is formed due to the electronegativity of the discharge. In an electronegative discharge, an `electropositive edge' is formed, where the electron density has a local maximum.  Therefore, there will be specific regions inside the plasma where the sign of the ambipolar electric field is opposite to that near the closest electrode (i.e. there will be a spatial region with positive ambipolar field near the powered electrode). This will allow a high temperature region during sheath collapse, as the electrons flowing towards the electrode will be accelerated by this ambipolar field, thus contributing to a temporally more symmetric electron temperature distribution. We also show that the Ohmic  power absorption on time average increases as a function of pressure. 

The paper is structured as follows: in section \ref{sec2},
the theoretical background of the analysis of electron power
absorption based on the Boltzmann equation is briefly explained. Section \ref{sec3} contains the description of the PIC/MCC simulation used to obtain input parameters for the method. In section \ref{sec4}, results are presented and discussed. Finally, in section \ref{sec5} conclusions are drawn.

\section{Theoretical background}\label{sec2}

 Our analysis is based on the first velocity moment equation of the 1D Boltzmann equation \cite{lafleur14}, i.e. the momentum balance equation: 

\begin{align}\label{boltz}
\frac{\partial}{\partial t}(mnu)+\frac{\partial}{\partial x}(mnu^2)=-enE-\frac{\partial p_{xx}}{\partial x} -\Pi_{\rm c}.
\end{align}
Here, $n$ and $u$ are the electron density and mean velocity, respectively, $m$ is the electron mass and $e$ the electron charge. $p_{xx}=mn(\langle v_x^2\rangle -u^2)$ denotes the diagonal element of the pressure tensor, where $v_x$ is the velocity of an individual electron in the $x$-direction, perpendicular to the electrodes. $\Pi_{\rm c}$ denotes the change of momentum due to collisions.
From this equation the total electric field can be expressed as a sum of different space- and time-dependent terms. To reduce the complexity of the analysis, the seven different terms used previously in \cite{schulze18} have been changed to three. In that sense this paper follows \cite{lafleur14} more closely. The three terms are as follows 

 \begin{align}\label{Eterm}
E_{\rm in}(x,t)&=-\frac{m}{n(x,t)e}\left[\frac{\partial}{\partial t}(n(x,t)u(x,t))+\frac{\partial}{\partial x}(n(x,t)u(x,t)^2)\right],\nonumber \\
E_{\rm \nabla p}(x,t)&= - \frac{1}{n(x,t)e} \frac{\partial}{\partial x} p_{xx}(x,t), \nonumber \\	
E_{\rm Ohm}(x,t)&=-\frac{\Pi_{\rm c}(x,t)}{n(x,t)e}. \nonumber\\
\end{align} 
The electric field term originating from the pressure gradient, $E_{\nabla p}$ is commonly split into two seperate parts,  $E_{\nabla n}$ and $E_{\nabla T}$, which have the following forms \cite{schulze18}:
\begin{align}\label{Egradp}
E_{\nabla n}(x,t)&=-\frac{T_{xx}(x,t)}{n(x,t)e}\frac{\partial n(x,t)}{\partial x}, \nonumber \\
E_{\nabla T}(x,t)&=-\frac{1}{e}\frac{\partial T_{xx}(x,t)}{\partial x}.
\end{align} 
Here $T_{xx}(x,t)$ denotes the electron temperature (measured in eV, called `temperature' in the following) which is given by the ideal gas law as $T_{xx}(x,t)=p_{xx}(x,t)/n(x,t)$. Each term in equation (\ref{Eterm}) has a distinct physical origin.  The inertia term, $E_{\rm in}$, is the electric field needed to balance the change in electron momentum. $E_{\nabla n}$ and $E_{\nabla T}$ are related to the pressure gradient in equation (\ref{boltz}).  
$E_{\nabla n}$ balances the force resulting from the electron density gradient and is, in quasineutral regions, identical to the `classical' ambipolar electric field (see e.g. \cite{schulze15}). $E_{\nabla T}$ balances the force due to the gradient of the electron temperature. This term disappears when an isothermal situation is assumed \cite{ChabertBook}.  We also allow a time dependence for the electron temperature, which is essential for a nonzero ambipolar electron power absorption on time average, as shown in \cite{schulze18}. $E_{\rm Ohm}$ is a result of collisions and incorporates the classical Ohmic electric field. 
Based on these electric field terms and the electron conduction current density, $j_{\rm e}(x,t)$, the total power absorbed by the electrons, $P_{\rm tot}(x,t)$ is calculated as 

\begin{equation}\label{Ptot}
P_{\rm tot}(x,t)=j_{\rm e}(x,t)E_{\rm tot}(x,t)=\sum_{i=1}^3P_i(x,t)=\sum_{i=1}^3j_{\rm e}(x,t)E_i(x,t).
\end{equation} 
The electric field terms and the power absorption terms (P$_i$-s) are calculated via equations (\ref{Eterm}) and (\ref{Ptot}), where the space- and time-dependent physical quantities ($n$, $u$, $p_{xx}$, $T_{xx}$, $\Pi_{\rm c}$, and $j_{\rm e}$) are taken from PIC/MCC-simulations. As no a priori assumptions have been made, the above described method is exact and accounts for every physical mechanism of electron power absorption. As the PIC/MCC simulations are self-consistent, the results are exact within the validity domain of our discharge model (which assumes a 1D geometry and neglects electromagnetic effects). The consistency and completeness of the method is checked in every scenario studied by comparing the sum of the electric field terms with the total electric field obtained directly from PIC/MCC-simulations. For more details see \cite{schulze18}.

\section{Computational method}\label{sec3}
The numerical calculations are based on our 1d3v PIC/Monte Carlo collision simulation code \cite{donkPIC}.
The processes taken into account in our oxygen discharge model are partially based on the `xpdp1' set of elementary processes \cite{vahedi} and its recent revision \cite{gud}. The charged species considered in the model are ${\rm O_2^+}$ and
${\rm O^-}$ ions and electrons. The set of elementary collision processes between the electrons and ${\rm O_2}$ neutral molecules
includes elastic scattering, excitation to rotational, vibrational and electronic levels, ionisation, dissociative excitation, dissociative attachment, impact detachment and dissociative recombination. For ${\rm O_2^+}$ ions, elastic collisions with ${\rm O_2}$ are taken into account; we include the symmetric charge exchange process and an additional channel with isotropic scattering in the centre-of-mass frame as suggested by \cite{gud}. For ${\rm O_2^-}$ ions, the model includes elastic scattering with ${\rm O_2}$ neutrals, detachment in
collisions with electrons and ${\rm O_2}$ molecules, mutual neutralisation with ${\rm O_2^+}$ ions, as well as collisions with metastable
singlet delta oxygen molecules, ${\rm O_2(a^1\Delta)}$, where the latter is known to play an important role in oxygen CCPs \cite{greb1,greb2,proto}. For further details see \cite{aranka, ang}. 

In the simulations, we assume plane and parallel electrodes with a gap
of $L = 25$ mm. One of the electrodes is driven by a voltage waveform $\phi=\phi_0\cos(2\pi ft)$ with $f=27.12$ MHz, $\phi_0=200$ V, while the other electrode is at ground potential. To simplify the analysis, secondary electron emission is omitted in the model, only the reflection of the
impinging electrons is taken into account with a probability of 20\% \cite{kollath}. The gas temperature is fixed at
$T_{\rm g}=350$ K.  For the surface quenching probability of ${\rm O_2(a^1\Delta)}$ singlet delta molecules we use the value of $\alpha=6\cdot10^{-3}$ \cite{aranka,ang,derzsi}. The computations are carried out using a spatial grid with $N_x=100-1600$ points and $N_t=2000-85000$ time steps within the RF period. These parameters have been set to fulfil the relevant stability criteria of the numerical method and to provide high resolution data. 
The computation of the different terms is incorporated into the code as given by equations (\ref{Eterm}) and (\ref{Ptot}). Due to the complexity of the model and hence the rapid increase of computational time with the increase of the number of  superparticles, to obtain high precision data we use  $\sim 10^5$ particles per species (${\rm O_2^+}$, ${\rm O^-}$ ions and electrons) and obtain data during $\sim10^4$ RF-cycles after full convergence with high spatial and temporal resolution. 

\pagebreak

\section{Results}\label{sec4}
Results are presented for a pressure range between 2 and 50 Pa for a driving voltage amplitude $\phi_0=200$ V and frequency $f=27.12$ MHz. The structure of presentation is as follows: First, the spatiotemporal density profiles of the two extremal cases (2 Pa and 50 Pa) are presented and analysed. Then results for the spatial distribution of the  time-averaged electron power absorption,  $P_{\rm tot}$, are presented, and, subsequently, each term in equations (\ref{Eterm}) and (\ref{Ptot}) is discussed for different pressures using the space- and time-resolved analysis described in section \ref{sec2}. 

\begin{figure}[H]
\centering
\begin{center}
\includegraphics[width=0.9\textwidth]{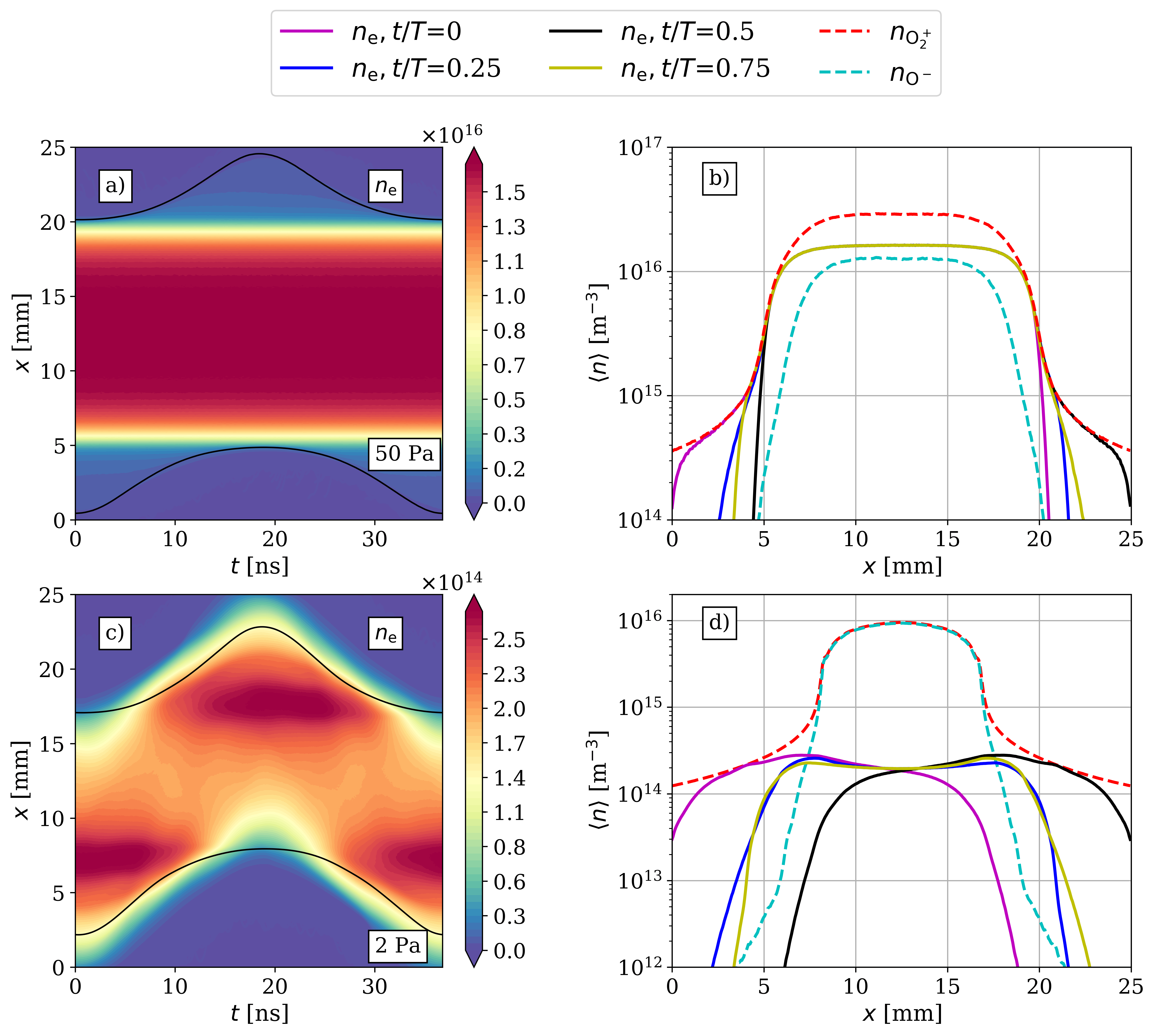} 
\caption{Spatiotemporal distribution of the electron density at 50 Pa (a) and 2 Pa (c) and the time-averaged ion densities together with the electron density at times $t/T=0$, 0.25, 0.5, 0.75 for 50 Pa (b) and 2 Pa (d). The black lines in panels (a) and (c) indicate the sheath edges calculated by the Brinkmann-criterion \cite{Brinkmann}. The powered electrode is situated at $x=0$, while the grounded electrode is at $x=25$ mm. $L=25$ mm, $\phi_0=200$ V, $f=27.12$ MHz. }
\label{n}
\end{center}
\end{figure}

To aid the understanding of the results to be presented, first the main characteristics of the electron density have to be analysed. Figure \ref{n} shows the spatio-temporal electron density profile as well as the electron density profile at times $t/T=0$, 0.25, 0.5, 0.75, together with the time averaged ion density profiles at 50 Pa (panels (a) and (b), respectively) and 2 Pa (panels (c) and (d)). As shown in \cite{ang}, in a low-pressure oxygen CCP driven by a single-frequency voltage waveform, the electronegativity of the plasma is very low at high pressures, and it increases with decreasing pressure. Due to this phenomenon, oxygen at high pressures when driven by a single frequency voltage waveform, exhibits `electropositive-like' properties \cite{gud2019}. This can be seen in panels (a) and (b): the electron density increases monotonically as a function of the distance from the electrode.

At low pressures the density profile shows a completely different behaviour. This is a consequence of the high electronegativity of oxygen discharges at low pressures. Because of the high electronegativity an `electropositive edge' is formed at the position of the maximum sheath edge  (see Figure \ref{n} (c), (d)), accounting for a local maximum of the electron density at that position \cite{schulze15}.  This local density maximum is formed because the negative ions inside the bulk have a relatively high mass, which leads to their `confinement' in the bulk region. As the presence of negative ions leads to the depletion of the electron density because of quasineturality, the electron density will be maximal at the edge of this `negative ion region'.

\begin{figure}[H]
\centering
\begin{center}
\includegraphics[width=0.9\textwidth]{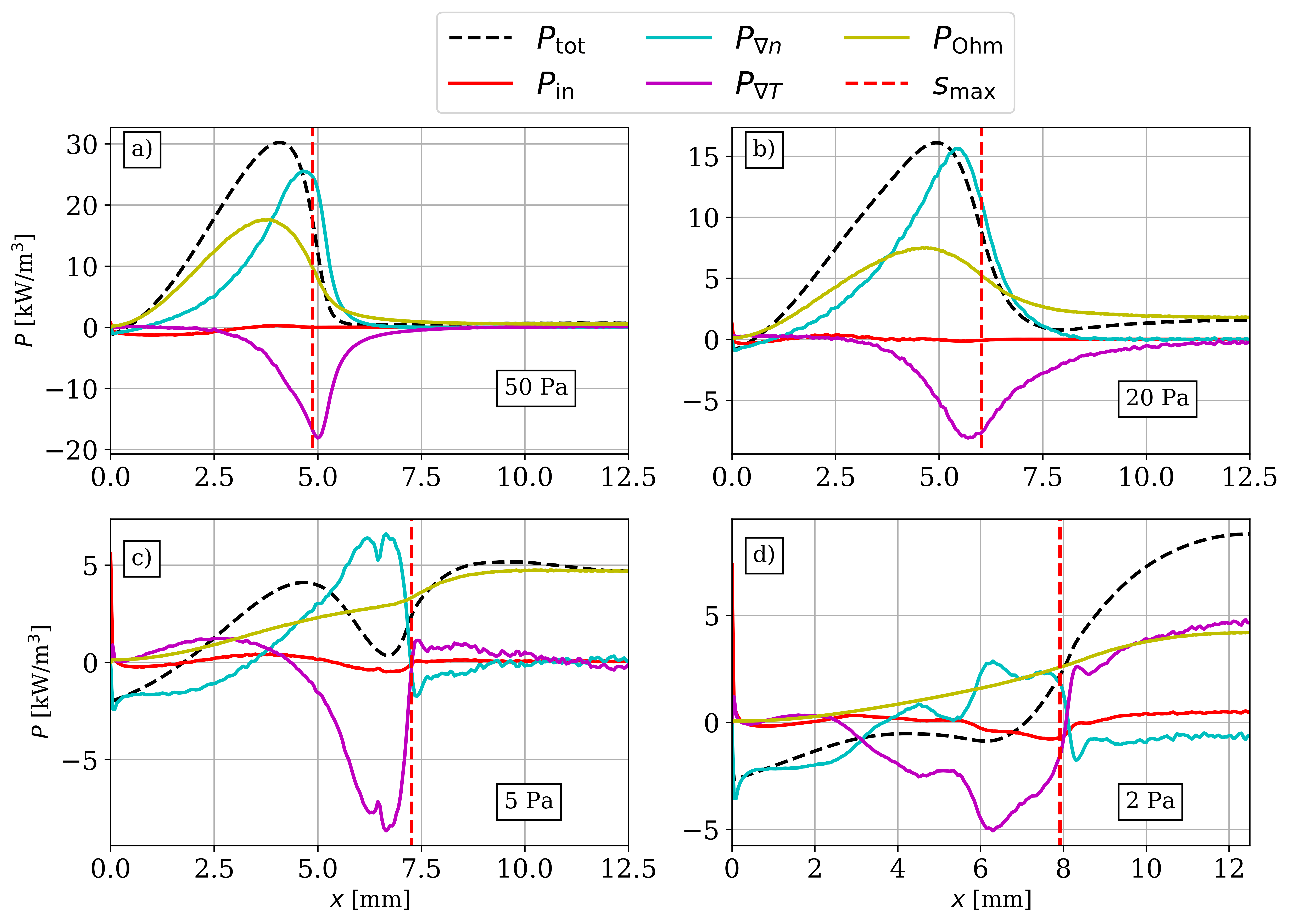} 
\caption{Spatial distribution of the time-averaged total electron power absorption, $P(x)$, and the contribution of individual terms, as defined in equation (\ref{Eterm}) at different pressures. (a) 50 Pa, (b) 20 Pa, (c) 5 Pa, and (d) 2 Pa. The plots show only the proximity of the powered electrode situated at $x=0$ mm. The electrode separation is $L=25$ mm. The vertical dashed lines indicate the maximum sheath width, $s_{\rm max}$. $\phi_0=200$ V, $f=27.12$ MHz.}
\label{timeav_power_dens}
\end{center}
\end{figure}

Figure \ref{timeav_power_dens} shows the time average of the total power absorbed by the electrons ($ P_{\rm tot}(x)=\frac{1}{T}\int\limits_0^TP_{\rm tot}(x,t)dt$) and the individual power absorption terms as given in equation (\ref{Ptot}) ($ P_i(x)$, $i=1..4$) as a function of distance from the powered electrode in a region adjacent to this electrode for different values of the gas pressure, i.e. for 50 Pa, 20 Pa, 5 Pa and 2 Pa (panels (a), (b), (c), (d), respectively). The sum of the values obtained for these quantities is in perfect agreement with the total electron power absorption obtained directly from the simulations, thus providing a consistency check of our model. The maximum length of the sheath, $s_{\rm max}$, is also included. Due to symmetry reasons the grounded electrode and its proximity is not shown.

It is instructive to start the analysis at the highest pressure, i.e. 50 Pa (panel (a)), because it appears to show the most simple structure of the four cases. In this case $P_{\nabla n}$, $P_{\nabla T}$ and $P_{\rm Ohm}$ dominate. $P_{\nabla n}$, which is related to the ambipolar electric field, is proportional to the electron density gradient (see equation (\ref{Eterm})). Therefore, it exhibits a maximum close to the position of maximal sheath width, $s_{\rm max}$, where the normalized electron density gradient, $\frac{1}{n}\frac{\partial n}{\partial x}$ is high (see Figure  \ref{timeav_power_dens}). This term is positive over the whole spatial domain, except for the immediate vicinity of the electrode, which originates from the fact, that during sheath collapse there is a small `floating potential' present to prevent excess electron loss to the electrodes \cite{schulze18}. $P_{\nabla T}$, being proportional to the electron temperature gradient, accounts for significant electron power loss on time average, partially compensating $P_{\nabla n}$ (as noted earlier, both terms are related to the pressure gradient in the momentum balance equation, see equation (\ref{boltz})). It exhibits a minimum around the position of the maximal sheath width, which is related to the formation of a double layer of charges (see later). The collisional power absorption term ($P_{\rm Ohm}$), which represents Ohmic power absorption, acquires its highest value in the sheath region and remains constant in the bulk domain, because the electron density and thus the conductivity of the electron is lowest in the sheath region (see also Figure \ref{n} (a)), as $\sigma\propto\omega_{\rm p,e}^2\propto n_{\rm e}$, where $\omega_{\rm p,e}$ is the electron plasma frequency \cite{MakabeBook}. As the pressure is decreased to 20 Pa (Figure \ref{timeav_power_dens}(b)), $P_{\nabla n}$ and $P_{\nabla T}$ exhibit the same properties as before, whereas $P_{\rm Ohm}$ decreases, as with the decrease of pressure, the conductivity increases (or equivalently, the collisionality decreases).

At lower pressures (5 Pa and 2 Pa, (c), (d) panels), rather counterintuitively, $P_{\rm Ohm}$ has a significant contribution to the time averaged total electron power absorption. $P_{\nabla n}$, the ambipolar power absorption gets attenuated, its contribution to the total electron power absorption decreases with decreasing pressure. $P_{\nabla T}$ increases in the bulk domain, becoming positive in the bulk domain for 2 Pa, thus accounting for strong bulk power absorption in that case. We note that the time averaged contributions of the inertia term, $P_{\rm in}$ is negligible in all cases considered.

 The physical origins of these phenomena can be understood only via the space and time resolved analysis of each term in equations (\ref{Eterm}) (electric field) and (\ref{Ptot}) (electron power absorption). Therefore, we compare the two extremal cases, i.e. 50 Pa and 2 Pa. 
 
\begin{figure}[H]
\centering
\begin{center}
\includegraphics[width=0.95\textwidth,height=0.95\textheight,keepaspectratio]{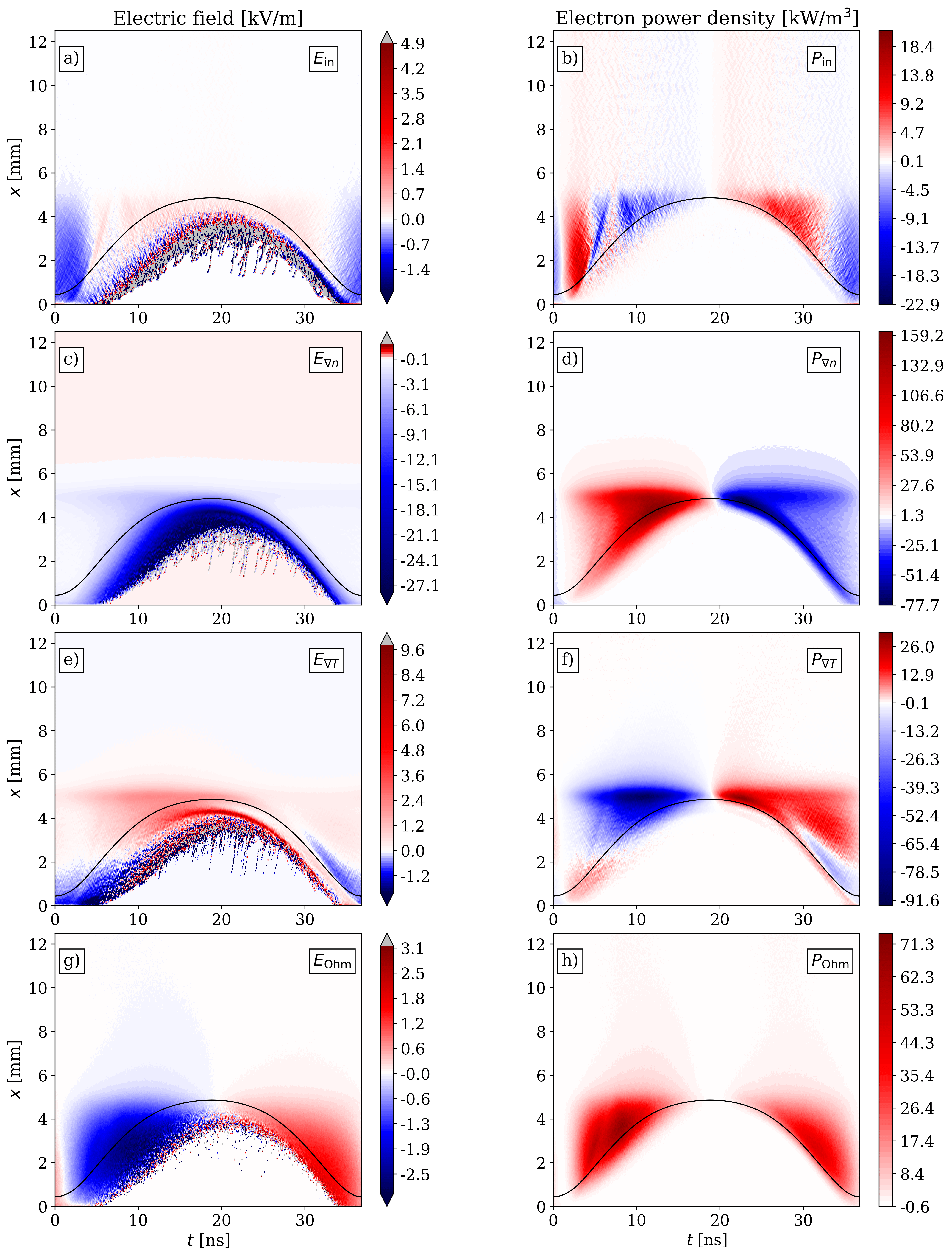} 
\caption{Spatio-temporal distribution of the electric field terms (left) and the electron power density terms (right) in the vicinity of the powered electrode
during one RF period at 50 Pa. The black line indicates the sheath edge.  The `noisy' features (lines, dots) in the electric field terms deep inside the sheath region originate from rare events of electron detachment processes, thus these represent real physical effects. $L=25$ mm, $\phi_0=200$ V, $f=27.12$ MHz.}
\label{50paall}
\end{center}
\end{figure}

\begin{figure}[H]
\centering
\begin{center}
\includegraphics[width=0.95\textwidth,height=0.95\textheight,keepaspectratio]{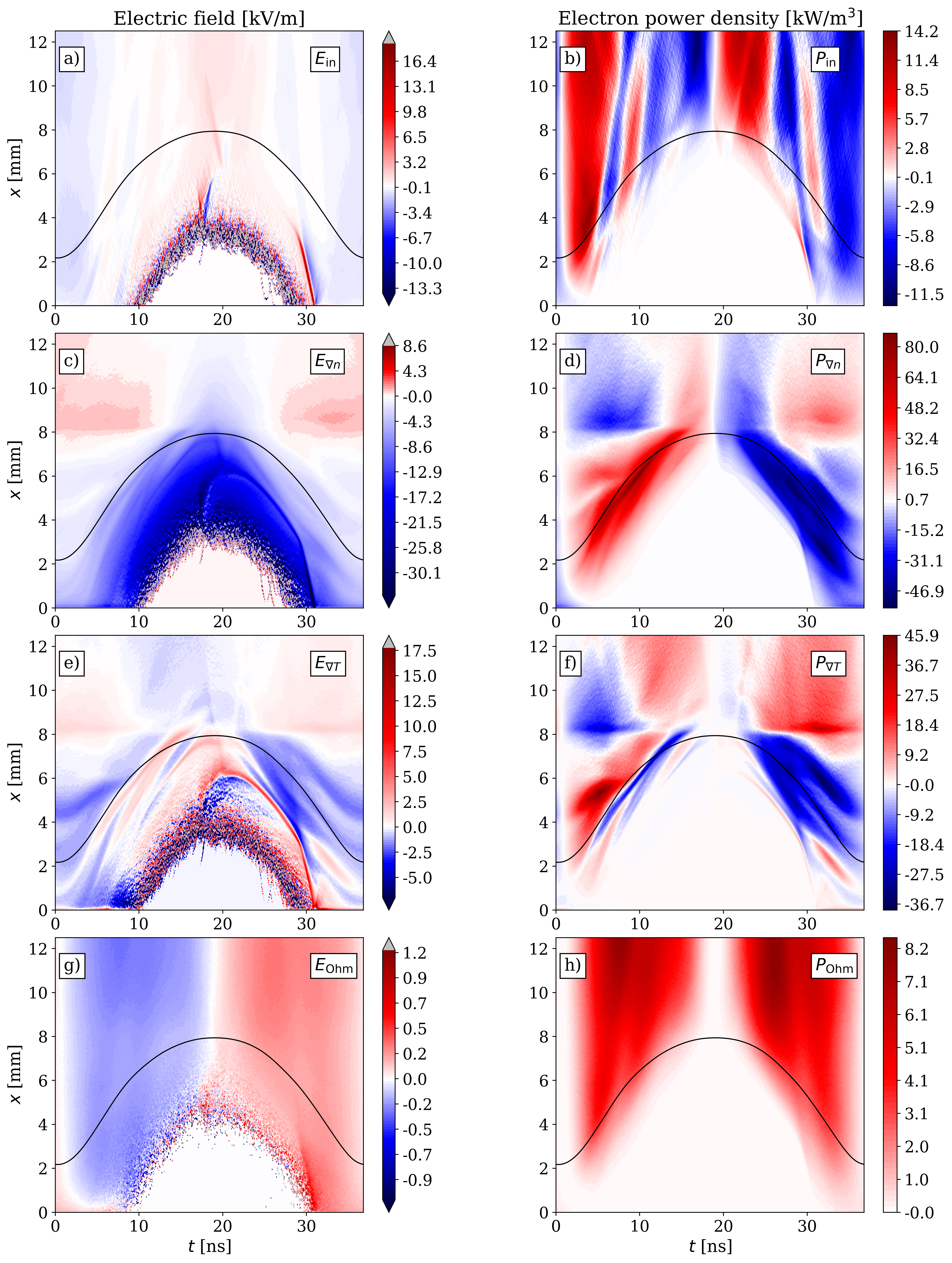} 
\caption{Spatio-temporal distribution of the electric field terms (left) and the electron power density terms (right) in the vicinity of the powered electrode
during one RF period at 2 Pa. The black line indicates the sheath edge.  The `noisy' features (lines, dots) in the electric field terms deep inside the sheath region originate from rare events of electron detachment processes, thus these represent real physical effects. $L=25$ mm, $\phi_0=200$ V, $f=27.12$ MHz.}
\label{2pall}
\end{center}
\end{figure}

 The space and time resolved analysis of the electric field and electron power absorption terms are shown in Figure \ref{50paall} in case of 50 Pa, and in Figure \ref{2pall} in case of 2 Pa. In both of these figures, the left column shows the four electric field terms, the right column the corresponding power absorption terms. These figures provide a complete description of the electron power absorption dynamics \cite{schulze18}.

We start with the highest pressure, i.e. 50 Pa. As shown in Figure \ref{n}, in a low-pressure oxygen CCP driven by a single-frequency voltage waveform, the electronegativity of the plasma is very low at high pressures (as in 50 Pa and also 20 Pa), and it increases with decreasing pressure \cite{ang}. Therefore, at this pressure we observe an `electropositive-like' behaviour. Because of this, the same charateristics are observed as in argon, for which a thorough analysis was presented in \cite{schulze18}. Therefore, in this paper we only give a brief summary of the high pressure case, emphasizing the most important characteristics which are different in the low-pressure, electronegative case.

The inertia term (see Figure \ref{50paall} (a)), $E_{\rm in}$, acquires its highest values in the vicinity of  the instantaneous sheath edge (where the electron density is low, see Figure \ref{n} (a), (b)), but its overall contribution to the total electric field and power absorption is negligible.  The strongest electric field terms close to the instantaneous sheath edge are  $E_{\nabla n}$ and $E_{\nabla T}$. These generate strong power absorption terms, too. It is important to note, that $E_{\nabla n}$ does not change its sign during the RF period (e.g. it is negative at the powered electrode). This is a consequence of the electron density profile at 50 Pa. As $E_{\nabla n}\propto-\frac{\partial n}{\partial x}$, and the electron density monotonically increases from the position of the electrode to the middle of the bulk, the gradient of the electron density will have the same sign.  $P_{\nabla n}$ exhibits two maxima, one near the instantaneous sheath edge, and one at the position of maximal sheath width. $P_{\nabla T}$, which is related to the elecron temperature gradient, exhibits a complex structure, which is due to the space- and time dependence of the electron temperature (see later). Although the contribution of the ohmic term, $E_{\rm Ohm}$, to the total electric field near the instantaneous sheath edge is small, it penetrates into the bulk region. $P_{\rm Ohm}$ has always positive values, since $j_{\rm e}$ and  $E_{\nabla n}$ always have the same sign, as they change directions simultaneously within the RF period. 

As the most important electric field term is $E_{\nabla n}$, one has to understand why it leads to a nonzero power absorption on time average. It has the same sign during sheath expansion as well as sheath collapse, since  $E_{\nabla n}$ is determined by the ions, which cannot react to the RF excitation. The electron current density changes sign during the sheath expansion/collapse phase \cite{schulze18}. Thus, without a temporal asymmetry of $T_{xx}$ within the RF period, there would be no ambipolar power absorption on time average.
 
\begin{figure}[H]
\centering
\begin{center}
\includegraphics[width=0.9\textwidth]{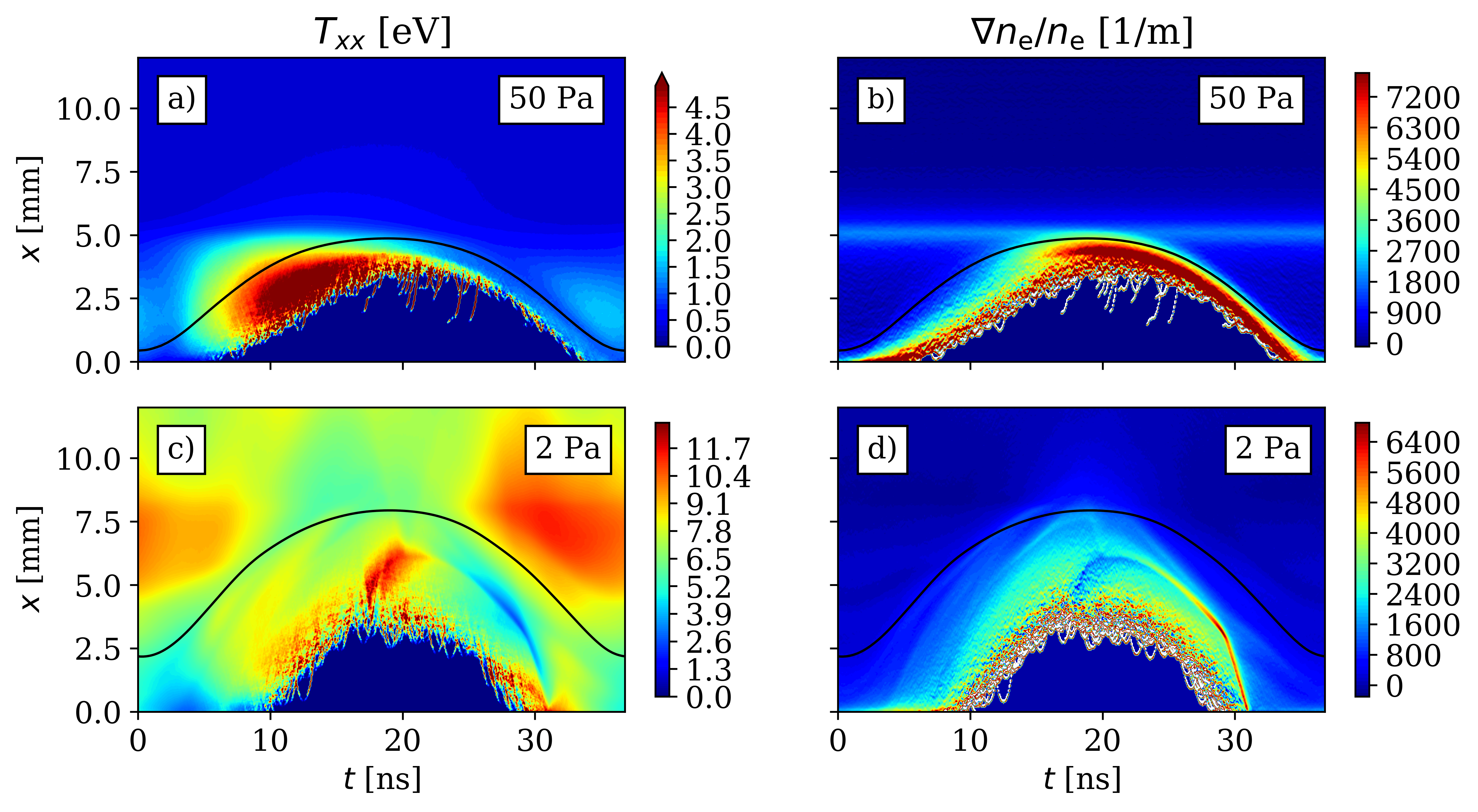} 
\caption{Spatio-temporal distribution of the electron temperature (left coloumn) and the normalized electron density gradient (right coloumn) in the vicinity of the powered
electrode during one RF cycle at 50 Pa (top row) and 2 Pa (bottom row). The black lines indicate the sheath edges.  The `noisy' features (lines, dots) deep inside the sheath region originate from rare events of electron detachment processes, thus these represent real physical effects.}
\label{density}
\end{center}
\end{figure}

The temporal asymmetry of the ambipolar field can be understood in more detail based on panels (a) and (b) of Figure \ref{density}, which show the electron temperature and the normalized electron density gradient at 50 Pa, respectively. The temporal modulation of the electron temperature is strong, as it is much higher in the sheath region of the expanding sheath compared to the sheath collapse phase, whereas the normalized electron density gradient does not change significantly during the two phases.
 
 The time modulation of the electron temperature is caused by energetic electrons generated by the electric field near the instantaneous sheath edge during sheath expansion, but not during sheath collapse. The most dominant term in the electric field is the ambipolar field itself. When the sheath starts to expand, cold electrons are accelerated towards the bulk. This increases the electron temperature, which increases the ambipolar field itself, thus leading to a self-amplifying mechanism. This mechanism stops at the maximum of the sheath edge (see Figure \ref{50paall} (c)). This does not work during sheath collapse, beacuse of cold electrons flowing towards the powered electrode, and also because of the high pressure, which, through collisions, prevents energetic electrons generated at the grounded electrode to reach the other electrode. As the ambipolar electric field has the same sign during both sheath expansion and sheath collapse, but the sign of the electron conduction current is different, the ambipolar field has a `cooling' effect on the electrons during sheath collapse, but on time average, this asymmetry results in a strong, positive ambipolar power absorption.  
 
 $E_{\nabla T}$, which is proportional to the electron temperature gradient, causes significant electron power loss on time average around the position of the maximum sheath edge during sheath expansion (Figure \ref{timeav_power_dens} (a), \ref{50paall} (f)). The reason for this `cooling' mechanism is that during sheath expansion   around the position of maximum sheath width, energetic electrons move towards a region of lower electron temperature. This gerenates a double layer of positive and negative space charges, the negative charges being at the bulk side. This, in turn, generates an electric field, which decelerates the electrons and results in a significant electron power loss.

Next, we analyse the behaviour of the electric field and power absorption terms at 2 Pa. Figure \ref{2pall} shows much more complex spatio-temporal distributions of the electric field as well as of the electron power absorption than in case of 50 Pa. In this case the inertial term, $E_{\rm in}$, is not negligible, but has a minor contribution to the total electron power absorption on time average (see Figure \ref{timeav_power_dens}). Although the absolute value of the Ohmic electric field,  $E_{\rm Ohm}$, is small, as compared to the other terms, it causes stronger bulk power absorption on time average, than in case of 50 Pa. This is caused by the decrease of electron density inside the bulk, which leads to a high electric field, as a consequence of the high electronegativity of oxygen at low pressures, and the presence of the `electropositive edge' shown in Figure \ref{n} (c),(d)). This has a very important effect on the ambipolar electric field (Figure \ref{2pall} (c)): Whereas at high pressures, the `electropositive-like' behaviour meant that the electron density increases monotonically as a function of distance from the powered electrode, thus accounting for a negative ambipolar electric field, in this case there will be two regions: The region between the electrode and the local maximum of the electron density, where the same conditions apply as at high pressure, and the region between the local maximum of the electron density and the inside of the bulk, where the exact opposite happens, i.e. the electron density decreases as a function of the distance from the adjecent electrode, thus resulting in a positive ambipolar electric field. 

Due to the electronegativity of the plasma, the term resulting from the gradient of the electron temperature ($E_{\nabla T}$, Figure \ref{2pall} (e)), shows a complex structure, which is present inside the bulk as well. The corresponding electron power absorption term (Figure \ref{2pall} (f)) shows a temporal asymmetry in the bulk region, which causes the significant bulk power absorption in this case.

\begin{figure}[H]
\centering
\begin{center}
\includegraphics[width=0.9\textwidth]{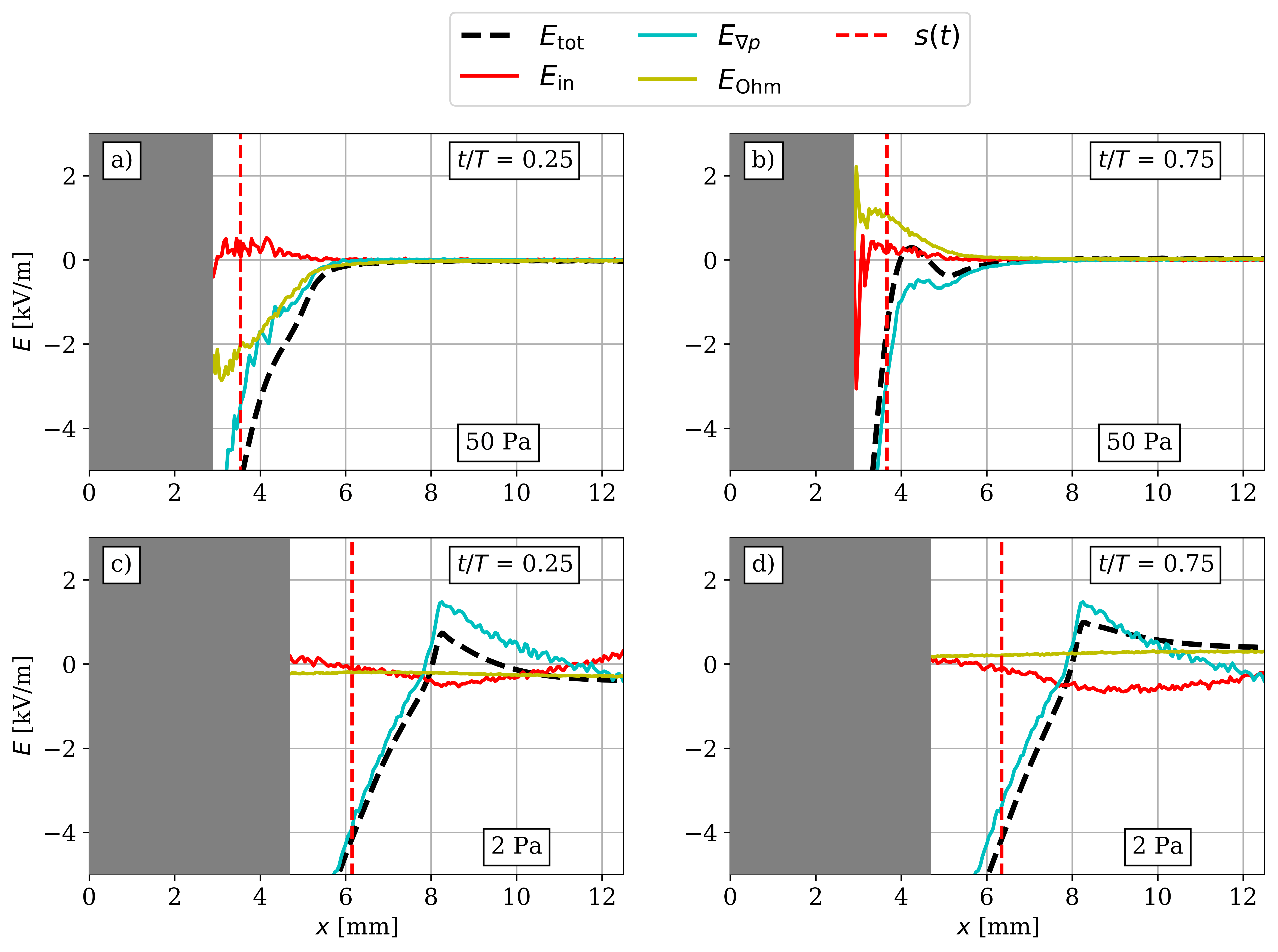} 
\caption{Spatial distribution of the electric field terms at times $t/T$ = 0.25 (a), (c) and $t/T$ = 0.75. The gas pressure is $p$ = 50 Pa (a), (b) and $p$ = 2 Pa (c), (d). The regions directly adjacent to the electrode are not shown, since the electron density in these regions vanishes and, thus, the signal to noise ratio of the simulation data is very low. The vertical dashed lines indicate the position of the instantaneous sheath edge. $L=25$ mm, $\phi_0=200$ V, $f=27.12$ MHz.}
\label{efieldmirror}
\end{center}
\end{figure}

\begin{figure}[H]
\centering
\begin{center}
\includegraphics[width=0.9\textwidth]{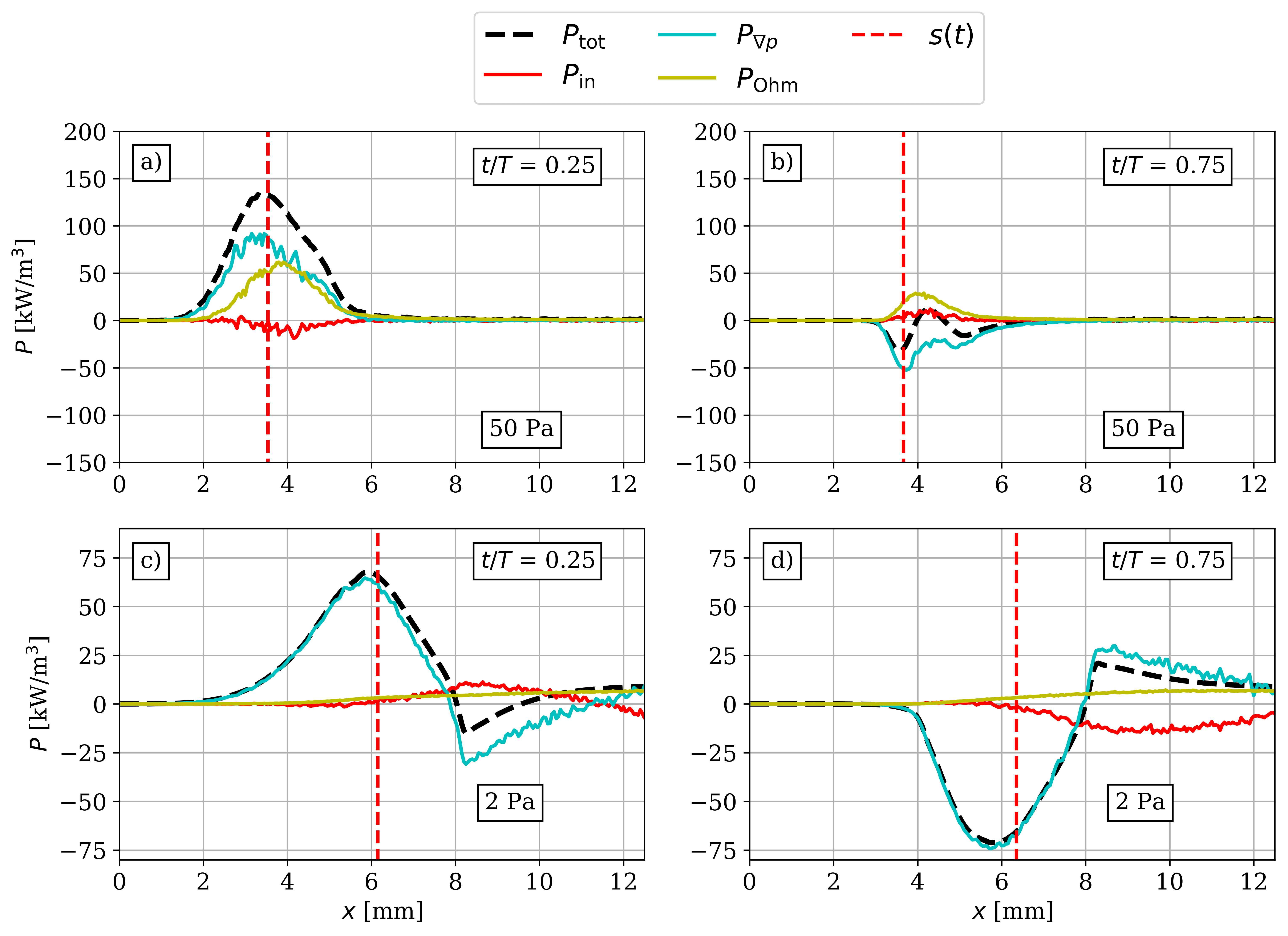} 
\caption{Spatial distribution of the electron power density terms at times $t/T$ = 0.25 (a), (c) and $t/T$ = 0.75 (b), (d). The gas pressure is
$p$ = 2 Pa (a), (b) and $p$ = 50 Pa (c), (d). The vertical dashed lines indicate the position of the instantaneous sheath edge. $L=25$ mm, $\phi_0=200$ V, $f=27.12$ MHz.}
\label{heatingmirror}
\end{center}
\end{figure}

Figures \ref{efieldmirror} and \ref{heatingmirror} show the spatial distributions of the electric field and electron power absorption terms at times $t/T$ = 0.25 and $t/T$ = 0.75 for 50 Pa ((a) and (b), respectively) and 2 Pa ((c) and (d), respectively). In case of 50 Pa we observe a strong, negative pressure gradient term (that is, the sum of $E_{\nabla n}$ and $E_{\nabla T}$), and a strong Ohmic electric field near the maximum of the sheath edge. The inertia term has a negligible contribution. In Figure  \ref{heatingmirror} (a) and (b) the temporal asymmetry of the electron power absorption can be observed: The power absorption terms in the two panels are not `mirror images' of each other, therefore leading to a nonzero electron power absorption on time average, as explained before.

The 2 Pa case shows a different behaviour: Figure \ref{efieldmirror} panels (c), (d) show that $E_{\nabla p}$ changes sign around the position of the maximal sheath edge. Furthermore, it increases as a function of distance from the powered electrode until the position of the local maximum of the electron density (it is zero at the maximum of the electron density), and decreases from the electropositive edge towards the bulk, as noted before. Although the inertial electric field term, $E_{\rm in}$, is significant, its contribution to the electron power absorption overall is negligible. The Ohmic power absorption term is small near the sheath edge and attains its maximal value inside the bulk region. The total electron power absorption in Figure \ref{heatingmirror} panels (c) and (d) is more symmetric than in the case of 50 Pa. As the most significant term in this low pressure case is $P_{\nabla p}$, which is intimately related to the ambipolar field, an explanation is needed as to why the ambipolar electric field is temporally more symmetric than at high pressure. This temporal symmetry of the ambipolar electron power absorption is what causes its attenuation on time average, thus leading to a significant contribution of the Ohmic power absorption term despite the low pressure.
 
To understand the temporal symmetry of the ambipolar electric field, we need the spatio-temporal distribution of the electron temperature and the normalized electron density gradient, as in the case of 50 Pa. Figure \ref{density} panels (c) and (d) show these quantities at 2 Pa, respectively.

The normalized density gradient is again symmetric as before. In the spatio-temporal distribution of the electron temperature, unlike the high presure case, regions with high electron temperature can be observed both in the expanding phase and in the collapsing phase. These high temperature regions are due to the electronegativity of the plasma. During sheath expansion, the same mechanism works as at high pressures, that is, the ambipolar electric field near the instantaneous sheath edge accelerates the electrons towards the bulk. At the position of the maximum sheath edge however, due to the electropositive edge, there is a positive ambipolar electric field,  because from the electropositive edge into the bulk the electron density decreases as a function of distance, therefore $-\frac{1}{n}\frac{\partial n}{\partial x}$ is positive, which decelerates the electrons, thus reducing their temperature. During sheath collapse, this positive ampipolar field will accelerate the incoming electrons, and the negative ambipolar field (generated between the powered electrode and the electropositive sheath edge) will decelerate them, therefore a region of high eletron temperature will form, farther from the powered electrode than during sheath expansion. This high temperature region will make the ambipolar electron power absorption more symmetric, thus leading to a reduced power absorption on time average. 

The attenuation of the cooling effect of the term proportional to the electron temperature gradient can be understood along the same reasoning: Whereas at high pressures a strong cooling was observed during sheath expansion due to the fact,  that a double layer of space charges formed at the position of the maximum sheath edge, with negative charges at the bulk side (or, equivalently, because the electron temperature is higher near the sheath than in the bulk, therefore $-\frac{\partial T}{\partial x}$ will be positive and this results in cooling), here, at low pressure the opposite effect is observed because of the positive ambipolar field at the electropositive edge, which reduces this cooling effect considerably. Thus, due to the temporal symmetry of the ambipolar electric field, the contribution of the term resulting from the electron temperature gradient to the total electron power absorption will be reduced on time average. Therefore, as  $P_{\nabla p}$ is the sum of these two terms, it will also be attenuated.

\begin{figure}[H]
\centering
\begin{center}
\includegraphics[width=0.9\textwidth]{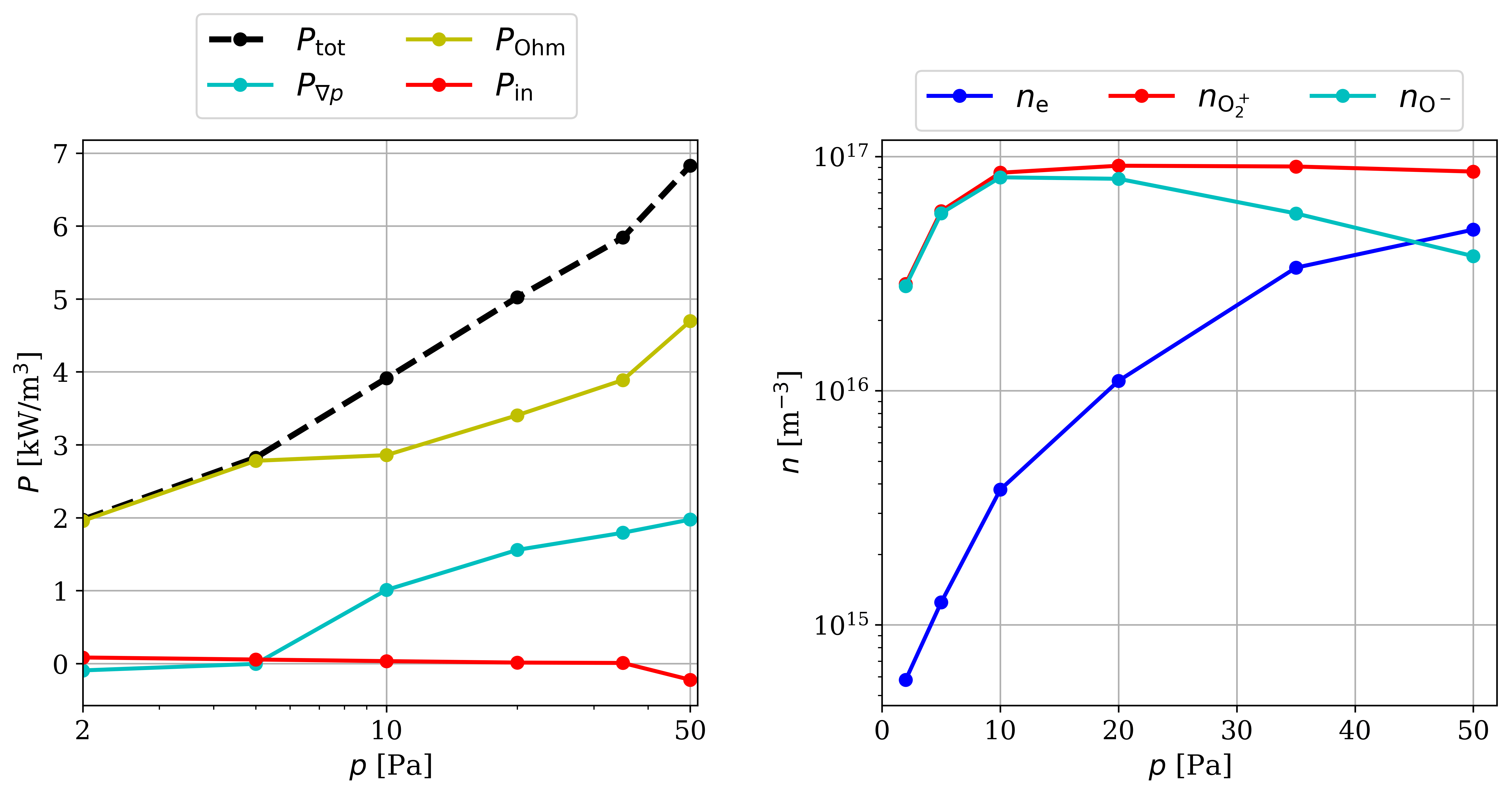} 
\caption{ Space- and time-averaged electron power density corresponding to each electric field term (a) and the space- and time averaged particle densities (b) as a function of pressure. $L=25$ mm, $\phi_0=200$ V, $f=27.12$ MHz. }
\label{spacetimeavg}
\end{center}
\end{figure}

Figure \ref{spacetimeavg} (a) shows the space- and time-averaged  absorbed electron power density corresponding to each term in equation (\ref{Ptot}), as well as the total absorbed electron power density 	as a function of pressure. $P_{\nabla p}$ is the sum of  $P_{\nabla n}$ and $P_{\nabla T}$, which together originate from the pressure gradient in equation (\ref{boltz}).  The inertial power absorption term, $P_{\rm in}$ is insignificant at all pressures. In all cases the most relevant terms are $P_{\rm Ohm}$  and $P_{\nabla p}$. The latter has a significant contribution to electron power absorption at high pressures, which decreases as the pressure is decreased, and vanishes completely at low pressures. The Ohmic term, $P_{\rm Ohm}$ is the most significant term throughout the whole pressure range, which monotonically increases as a function of pressure. Even at low pressures this term is the most significant, which is a deeply counterintuitive result, given the fact, that at low pressures the probability of collisions is reduced.

To understand, why the Ohmic power absorption term, $P_{\rm Ohm}$ is the most significant even at low pressures, where one would expect this term to be negligible, we need to look at Figure \ref{spacetimeavg} (b), which shows the time averaged particle densities in the middle of the plasma bulk as a function of pressure. This shows that the electron density at 2 Pa is two orders of magnitude lower than at high pressures. Therefore, the DC conductivity ($\sigma$) of the plasma significantly decreases towards lower pressures, as $\sigma\propto\omega_{\rm p,e}^2\propto n_{\rm e}$ \cite{MakabeBook}. Therefore, the plasma resistivity, being the inverse of the conductivity, significantly increases, which leads to an increase of the Ohmic power absorption term. But in order to explain why this term is the most significant, we also have to explain the attenuation of $P_{\nabla p}$, i.e. the term associated with the pressure gradient at low pressures, which was most significant in case of argon at low pressures \cite{schulze18}. At high pressures $P_{\nabla p}$ has a high positive value, which decreases when the pressure is lowered and vanishes at low pressures on time average. The reason for this is the attenuation of the ambipolar electric field, which is due to the pressure dependence of the electronegativity of oxygen, as oxygen is more electronegative at low pressures, thus having an electropositive edge, which leads to a positive ambipolar electric field (at the bulk side of the local maximum of the electron density in the electropositive edge region). This results in a temporally symmetric electron temperature distribution and thus an attenuation of $P_{\nabla p}$ on time average. 

\section{Conclusions}\label{sec5}

The electron power absorption dynamics in low pressure capacitive RF plasmas operated in oxygen at a frequency of 27.12 MHz were studied within the spatially and temporally resolved RF period based on the momentum balance equation derived from the Boltzmann equation using input parameters from 1d3v electrostatic PIC/MCC simulations. As this method does not make any a priori assumptions, it is fully self-consistent and exact as far as the PIC/MCC simulations allow, therefore it provides a deep understanding of this complex phenomenon. By invoking the momentum balance equation, the electric field and the absorbed electric power density of the electrons can be constructed as the sum of three different terms, respectively, each corresponding to a different physical mechanism. Through the analysis of the spatio-temporal distributions of these terms individually as a function of neutral gas pressure, the dominant electron power absorption mechanisms were identified. 

At high pressures (20-50 Pa), due to the low electronegativity of the oxygen discharge, similar results have been found as in argon \cite{schulze18}. The most significant part of the electron power absorption originates from the Ohmic term, which is due to collisions, and the ambipolar field, which is caused by the electron density gradient and a nonzero electron temperature. As noted in \cite{schulze18}, to have a nonzero power absorption generated by the ambipolar field on time average, a temporal modulation of the electron temperature within the RF-period is required. Furthermore, this modulation has to be temporally asymmetric as well, i.e. at a given spatial position $T_{xx}$ must have a different value during sheath expansion and sheath collapse. At higher pressures in oxygen, this occurs via a self-amplyfing mechanism: During sheath expansion, electrons are accelerated by the ambipolar electric field, which leads to an increase in the ambipolar electric field itself. This stops when the sheath is fully expanded.  The same mechanism does not apply during sheath collapse, as the sign of the electron current is reversed, whereas the sign of the ambipolar field remains the same. Thus, a temporally asymmetric electron temperature distribution forms, which allows the ambipolar field to generate a nonzero power absorption on time average.

 At low pressures (2 Pa), where oxygen is highly electronegative, it was found that $P_{\nabla p}$, i.e. the term associated with the pressure gradient is attenuated on time average, thus allowing the Ohmic power absorption term to be the most significant despite the low neutral gas pressure. This counterintuitive result can be explained by the electronegativity of the oxygen plasma, in which an electropositive edge region is formed with a local maximum of the electron density.  Therefore, the ambipolar field, which is proportional to the gradient of the electron density, will have a negative sign between the electrode and the local maximum of the electron density (where the electron density increases as a function of the distance from the powered electrode), and a positive part in the region between the local maximum and the bulk plasma. This will generate a temporally more symmetric scenario, which will strongly reduce the space- and time -averaged electron power absorption. During sheath expansion, the same mechanism is present as at high pressures, but during sheath collapse there will be another high temperature region, which is due to the positive ambipolar electric field, which will accelerate the incoming electrons. Overall, this results in the attenuation of both the ambipolar power absorption term, as well as the term proportional to the electron temperature gradient, and thus the term originating from the pressure gradient on time average.

\section{Acknowledgements}

This work was funded by the German Research Foundation in the frame of the project ,"Electron heating in capacitive RF plasmas based on moments of the Boltzmann equation: from fundamental understanding to knowledge based process control'' (No. 428942393). Support by the DFG via SFB TR 87, project C1, and by the Hungarian Office for Research, Development, and Innovation (NKFIH 119357) is gratefully acknowledged.

\end{document}